\documentclass{article}[12pt,a4paper]

\title{J/ $\Psi$ - J/$\Psi$ scattering cross sections of Quadratic and Cornell Potentials}
\usepackage[pdftex]{graphicx}
\usepackage{amsmath}
\usepackage{amssymb}
\usepackage{lmodern}
\usepackage{mathtools}
\usepackage{geometry}
\usepackage{tensor}
\usepackage{float}
\usepackage{subfigure}
\usepackage{bigints}
\usepackage{authblk}
\usepackage{caption}
\restylefloat{table}
\title{\textbf{J/ $\Psi$ - J/$\Psi$ scattering cross sections of Quadratic and Cornell Potentials}}
\author[1]{M. Imran Jamil}
\author[2]{S.M. Sohail Gilani}
\author[3]{Ahmad Wasif}
\author[4]{Abdul Sattar Khan}
\author[5]{Ahmad Awan}
\affil[1]{Department of Physics, School of Science, University of Management and Technology , Lahore (54770), Pakistan.}
\affil[2]{Department of Physics,University of Okara, Okara (56300), Pakistan.}
\affil[3]{Department of Physics, Forman Christian College (A Chartered University), Lahore (54600), Pakistan.}
\affil[4]{Department of Physics, Government College University, Lahore (54000), Pakistan.}
\affil[5]{Department of Physics, Government College University, Lahore (54000), Pakistan.}
\date{}
\begin{document}
\maketitle
\begin{abstract}

We study the scattering of J/$\Psi$ - J/$\Psi$ mesons using quadratic and Cornell potentials in our tetraquark (c\={c}c\={c}) system. The system’s wavefunction in the restricted gluonic basis is written by utilizing adiabatic approximation and Hamiltonian is used via quark potential model. Resonating group technique is used to get the integral equations which are solved to get the unknown inter-cluster dependence of the total wavefunction of our tetraquark system. T-Matrix elements are calculated from the solutions and eventually the scattering cross sections are obtained using the two potentials respectively. We compare these cross sections and find that the magnitude of scattering cross sections of quadratic potential are higher than Cornell potential.

\end{abstract}

\section{Introduction}

The $J/\Psi$ meson, a bound state of a charm quark and its anti-particle was discovered in 1974 and since then, it has been studied extensively from a theoretical and experimental point of view. In heavy ion collisions at RHIC \cite{PHENIX} and recently at LHC \cite{ALICE}, its production mechanisms have been explored. Perhaps, the most significant phenomenon as a result of heavy ion collisions is a formation of the particular state of matter called the quark-gluon plasma (QGP). A fundamental effect associated with the QGP medium is known as the color screening i.e. the interaction range of heavy quarks decreases with the increase in surrounding temperature \cite{QGP}. As a result, the potential between heavier quarks $c\bar{c}$ or $b\bar{b}$ gets screened due to the deconfinement of other quarks and gluons. The consequent separation of heavy quarks leads to suppressed quarkonia yields. J/ $\Psi$  suppression, an idea first put forward by theorists T. Matsui and H. Satz \cite{SUPPRESS} in 1986, is considered one of the indicators of the formation of QGP. Since charm quark is more abundantly produced in heavy-ion experiments as compared to bottom quark, researchers initially thought that J/ $\Psi$ could also be used for measuring the temperature of QGP. However, due to technical issues associated with $J/ \Psi$ production it is an unsuitable QGP temperature probe. Among J/ $\Psi$'s dominant decay modes is the production of lepton pairs. However, these dileptons do not possess the requisite momentum which can carry them beyond CMS particle spectrometer's large magnetic fields for detection. Furthermore, non-QGP effects can also play a role in the suppression of J/ $\Psi$. A better temperature probe which has been studied is the upsilon meson (bound state of bottom quark-anti quark) \cite{vogt}.
 \par
Various papers have considered the dissociation of $J/\Psi$ by light hadrons \cite{kharseev}-\cite{sibirtsev}. Due to the dominant scattering mechanism and the different assumptions made regarding it, the computed cross sections in these studies exhibit great variation at low energies. For example, Kharseev et.al \cite{kharseev},\cite{kharzsatz} studied the $J/\Psi$-nucleon collision using the parton model and perturbative QCD "short distance" approach. The cross section they obtained at $\sqrt{s} = 5$ GeV came out to be about 0.25 $\mu b$. Another study concerning dissociation cross sections of $J/\Psi$ by $\pi$ and $\rho$ mesons was done by Matinyan and Muller \cite{mullermat}. At $\sqrt{s} = 4$ GeV, they obtained $\sigma$ $\approx$ 0.2-0.3 $mb$. The dissociation cross sections mentioned above attain particular significance once one considers the suppression of quarkonium states (such as the stronger $\psi$(2$S$) suppression relative to the $J/\Psi$) due to the comover scattering effect \cite{gavin},\cite{capella}. Other theoretical studies have also probed various phenomenological aspects related to $J/\Psi$ meson.
\par
 Lattice QCD calculations done by \cite{m,p,q,r,s} found a relatively narrow width of J/ $\Psi$ i.e. around 1.6 $T_{c}$($T_{c}$ is known as the critical phase transition temperature). Afterwards, Wong \cite{dd} calculated J/ $\Psi$ production cross section via charm quark-anti quark collision and showed the energy dependance of cross section. Also, \cite{dd} concluded that with the corresponding decrease of temperature, an increase in maximum cross section occurs.
Theoretically, the two broad approaches used to study the dynamics of quarkonia are based on potential models and lattice QCD \cite{QR}.
\par
Potential models assume that the gluonic field energy between a heavy quark and its corresponding antiquark can be modelled with the aid of a suitable potential. For a single quark-antiquark pair, the Cornell potential adequately describes the experimental results of quarkonium spectroscopy \cite{agne1}, it also agrees well on the lattice \cite{agne2} and it can be obtained via quantum chromodynamics (QCD) \cite{agne3}. This quark-antiquark system can be treated in the non-relativistic regime owing to the heavier quark mass ($m_{c,b} \gg \Lambda_{\textsc{QCD}}$) and the low heavy quark velocity ($v \ll 1$). J. Weinstein and N. Isgur used a sum of two-body Cornell potentials in their study of K$\bar{K}$ molecule \cite{ISGUR,ISGUR2}. Afterwards, T. Barnes and E. Swanson employed the aforementioned approach \cite{SWA} to compute the cross sections and elastic scattering phase shifts of  $\pi^{+}\pi^{+}$, $K^{+}K^{+}$ and $\rho^{+}\rho^{+}$.
\par
Another potential that is of interest to our current study is the quadratic potential. It has been widely used to probe nucleon-nucleon interaction \cite{R1,R2,R3,R4} and recently to calculate the mass spectra of tetraquarks \cite{ZAH} . Quadratic confinement has also been used to investigate a six-quark state with a structure akin to benzene \cite{BEN}.
\par
In our study, we consider a system of two J/ $\Psi$ mesons and study their scattering cross sections with the aid of different potentials. We analyze how changing the potential between quarks would change the interaction between them. In Section 2, we write the wavefunction of our system followed by the Hamiltonian. The section ends with a discussion on quadratic and Cornell potentials. In Section 3, integral equations are written and decoupled using Born approximation. Our last section is devoted to results of quadratic and Cornell potentials.

\section{Wave function and the Hamiltonian}
We write the state function of our tetraquark system by utilizing adiabatic approximation \cite{EPJ} as follows
\begin{equation}
|\Psi_{T}\rangle = \sum_{k=1}^{2} \psi_{c}(\mathbf{Q}_{c})\chi_{k}(\mathbf{Q}_{k})\xi_{k}(\mathbf{u}_{k})\zeta_{k}(\mathbf{v}_{k})|k\rangle_{f}|k\rangle_{s}|k\rangle_{c}
\end{equation}
where, \\
$\mathbf{Q}_{c}$ : position of COM (center of mass) of our complete system, \\
 $\mathbf{Q}_{1}$ : vector joining COM of (1,$\bar{3}$) and (2,$\bar{4}$) clusters, \\
  $\mathbf{u}_{1}$ : vector representing position of quark 1 w.r.t. quark $\bar{3}$ in the cluster (1,$\bar{3}$), \\
  $\mathbf{v}_{1}$ : vector representing position of quark 2 w.r.t. quark $\bar{4}$ in the cluster (2,$\bar{4}$). \\
   Identical expressions are given for the clusters (1,$\bar{4}$) and (2,$\bar{3}$) via $\mathbf{Q}_{2}$, $\mathbf{u}_{2}$ and $\mathbf{v}_{2}$ . Likewise, (1,2) and ($\bar{3}$,$\bar{4}$) are defined by  $\mathbf{Q}_{3}$, $\mathbf{u}_{3}$ and $\mathbf{v}_{3}$. Hence,
\begin{eqnarray}\label{POS}
\mathbf{Q}_{1} = \frac{1}{2}(\mathbf{\tau}_{1} + \mathbf{\tau}_{\bar{3}} - \mathbf{\tau}_{2} -\mathbf{\tau}_{\bar{4}}) \nonumber \\
\mathbf{u}_{1} = \mathbf{\tau}_{1} - \mathbf{\tau}_{\bar{3}} \ , \  \mathbf{v}_{1} = \mathbf{\tau}_{2} - \mathbf{\tau}_{\bar{4}}
\end{eqnarray}
\begin{eqnarray}\label{POS1}
\mathbf{Q}_{2} = \frac{1}{2}(\mathbf{\tau}_{1} + \mathbf{\tau}_{\bar{4}} - \mathbf{\tau}_{2} -\mathbf{\tau}_{\bar{3}}) \nonumber \\
\mathbf{u}_{2} = \mathbf{\tau}_{1} - \mathbf{\tau}_{\bar{4}} \ , \  \mathbf{v}_{2} = \mathbf{\tau}_{2} - \mathbf{\tau}_{\bar{3}}
\end{eqnarray}
\begin{eqnarray}\label{POS2}
\mathbf{Q}_{3} = \frac{1}{2}(\mathbf{\tau}_{1} + \mathbf{\tau}_{2} - \mathbf{\tau}_{\bar{3}} -\mathbf{\tau}_{\bar{4}}) \nonumber \\
\mathbf{u}_{3} = \mathbf{\tau}_{1} - \mathbf{\tau}_{2} \ , \  \mathbf{v}_{3} = \mathbf{\tau}_{\bar{3}} - \mathbf{\tau}_{\bar{4}}
\end{eqnarray}
\begin{eqnarray}\label{POS3}
\xi_{k}(\textbf{u}_{k}) = \frac{1}{(2\pi d^{2})^{3/4}}\textrm{exp}\Big(\frac{- \textbf{u}^{2}_{k}}{4d^{2}}\Big) \nonumber \\
\zeta_{k}(\textbf{v}_{k}) = \frac{1}{(2\pi d^{2})^{3/4}}\textrm{exp}\Big(\frac{- \textbf{v}^{2}_{k}}{4d^{2}}\Big)
\end{eqnarray}
The following diagrams show the possible  topologies of quark anti-quark clusters:
\begin{figure}[H]
\centering
  \includegraphics[scale=0.5]{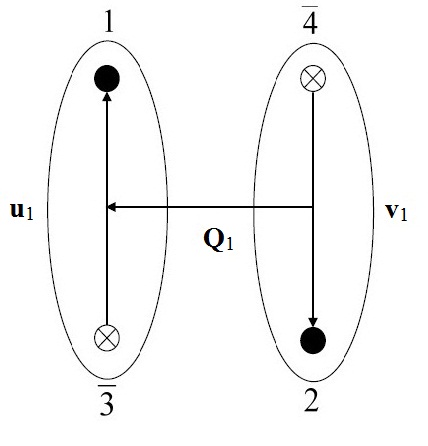}
  \caption{\textbf{Figure 1} \hspace{1mm} Topology A}
  \vspace{3mm}
  \includegraphics[scale=0.6]{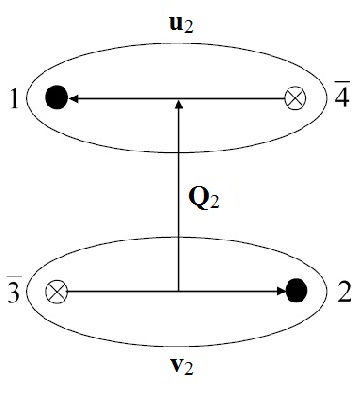}
  \caption{\textbf{Figure 2} \hspace{1mm} Topology B}
\end{figure}

As we are considering the system in centre of mass reference frame $\psi_{c}(\mathbf{Q}_{c})$ has no role in the dynamics of the four quark system. $\chi_{k}(\mathbf{Q}_{k})$ is an unknown in the radial part of total wavefunction and $\xi_{k}(\mathbf{u}_{k})$,$\zeta_{k}(\mathbf{v}_{k})$ are the predefined Gaussian wave functions. $|k\rangle_{f}$, $|k\rangle_{s}$ and $|k\rangle_{c}$ are flavor, spin and color parts of the wave function respectively.
The Hamiltonian of our (c\={c}c\={c}) system is written as follows:
\begin{equation}
\hat{H} =  \sum_{i=1}^{\bar{4}} \bigg[m_{i} + \frac{\hat{P}_{i}^{2}}{2m_{i}}\bigg] + \sum_{i<j}^{\bar{4}} [v(r_{ij})\mathbf{F_{i}}.\mathbf{F_{j}}]
\end{equation}
where $m$ and $\hat{P}$ denote quark mass and linear momentum respectively. Also, $F_{i} = \lambda_{i}/2$ for quark and $F_{i} = - \lambda_{i}^{*}/2$ for an anti-quark, where $\lambda$'s are the well-known Gell-Mann matrices. The potentials $v(r_{ij})$ used for $q\bar{q}$ pair wise interaction are Quadratic Potential \cite{MASUD}
\begin{equation}
v_{ij} = Cr^{2}_{ij} + \bar{C}
\end{equation}
where $i,j = 1,2,\bar{3},\bar{4}$
and Cornell Potential (Coulombic plus linear potential) \cite{ISGUR} is written as
\begin{equation}
v_{ij} = - \frac{4}{3} \frac{\alpha_{s}}{r_{ij}} + b_{s}r_{ij}
\end{equation}
with $i,j = 1,2,\bar{3},\bar{4}$
where
$\alpha_{s}$ is the strong coupling constant and $b_{s}$ is the string tension (Flux tube model)

The mesonic size $d$ appearing in $\xi_{k}(\mathbf{u}_{k})$ and $\zeta_{k}(\mathbf{v}_{k})$ can be adjusted in a manner such that the Gaussian ground state wave function of quadratic potential approximates the ground state wave function of Cornell potential. Hence, their overlaps become unity for a fitted value of parameter $d$ \cite{ACKLEH}.
\section{Integral equations and their solutions}
We employ the RGM (resonating group method) technique \cite{WHEELER} and take variations only in the $\chi_{k}$ $(\mathbf{Q}_{k})$ factor of the total wave function $\Psi_{T}$, and by making the use of linear independence of these variations both w.r.t. the two values of k, i.e. k=1,2 and w.r.t. all possible continuous values of $\mathbf{Q}_{k}$ in $\langle \delta \Psi_{T}| H - E_{c}|\Psi_{T}\rangle = 0 $  we get, from eq.1, the following two integral equations

\begin{equation}
\sum_{l=1}^{2} \int d^{3}u_{k} d^{3}v_{k} \xi_{k}(\mathbf{u}_{k})\zeta_{k}(\mathbf{v}_{k}) {}_{f} \langle k| {}_{s}\langle k| {}_{c}\langle k| \hat{H} - E_{c}|l\rangle_{c}|l\rangle_{s}|l\rangle {}_{f}\chi_{l}(\mathbf{Q}_{l})\xi_{l}(\mathbf{u}_{l})\zeta_{l}(\mathbf{v}_{l}) = 0
\end{equation}

The operator ($\hat{H}$ - $E_{c}$) is identity in the flavour basis whereas the overlap factors are also the ($\hat{H}$ - $E_{c}$) operator's flavour matrix elements. So, our potential energy matrix in spin and colour basis is
\begin{equation*}
V \equiv \langle k|\hat{V}|l\rangle_{cs} = {}_{s}\langle k|l\rangle_{s} {}_{c}\langle k|\hat{V}|l\rangle_{c}
\end{equation*}
 \[ =
\begin{bmatrix}
\ -\frac{4}{3}(v_{1\bar{3}} + v_{2\bar{4}}) &&&&&&&& \ -\frac{1}{2}\frac{4}{9}(v_{12} - v_{1\bar{3}} - v_{1\bar{4}} - v_{2\bar{3}} - v_{2\bar{4}} + v_{\bar{3}\bar{4}} \\
\ -\frac{1}{2}\frac{4}{9}(v_{12} - v_{1\bar{3}} - v_{1\bar{4}} - v_{2\bar{3}} - v_{2\bar{4}} + v_{\bar{3}\bar{4}} &&&&&&&& \ -\frac{4}{3}(v_{1\bar{4}} + v_{2\bar{3}})
\end{bmatrix},
\]
The spin overlaps are given as
\begin{equation}
{}_{0}\langle V_{1\bar{3}}V_{2\bar{4}}|V_{1\bar{4}}V_{2\bar{3}}\rangle_{0} = {}_{0}\langle V_{1\bar{4}}V_{2\bar{3}}|V_{1\bar{3}}V_{2\bar{4}}\rangle_{0} = -\frac{1}{2}
\end{equation}

where $V_{q\bar{q}}$ is a vector meson. Now, we discuss the Kinetic energy operator. In spin space $\hat{K}$ and  $\sum\limits_{i=1} ^{\bar{4}} [m_{i} - E_{c}]$ are unit operators. Thus, the matrix elements in spin-colour basis are similar to matrix elements in color basis multiplied by spin overlaps, i.e.
\begin{equation}
\langle k|\hat{K}|l\rangle_{cs} = {}_{s}\langle k|l\rangle_{s} {}_{c}\langle k|\hat{K}|l\rangle_{c}
\end{equation}
where k=1,2. Opening the summation over l, we have the following two equations respectively
\begin{eqnarray}\label{A}
\int d^{3}u_{1} d^{3}v_{1} \xi_{1}(\mathbf{u}_{1})\zeta_{1}(\mathbf{v}_{1}) {}_{c}\langle 1|\hat{H} - E_{c}|1\rangle_{c} \chi_{1}(\mathbf{Q}_{1})\xi_{1}(\mathbf{u}_{1})\zeta_{1}(\mathbf{v}_{1}) \nonumber \\
 + ( -\frac{1}{2}) \int d^{3}u_{1} d^{3}v_{1} \xi_{1}(\mathbf{u}_{1})\zeta_{1}(\mathbf{v}_{1}) {}_{c}\langle 1|\hat{H} - E_{c}|2\rangle_{c} \chi_{2}(\mathbf{Q}_{2})\xi_{2}(\mathbf{u}_{2})\zeta_{2}(\mathbf{v}_{2}) = 0
\end{eqnarray}
\begin{eqnarray}\label{B}
\int d^{3}u_{2} d^{3}v_{2} \xi_{2}(\mathbf{u}_{2})\zeta_{2}(\mathbf{v}_{2}) {}_{c}\langle 2|\hat{H} - E_{c}|1\rangle_{c} \chi_{1}(\mathbf{Q}_{1})\xi_{1}(\mathbf{u}_{1})\zeta_{1}(\mathbf{v}_{1}) \nonumber \\
 + ( -\frac{1}{2}) \int d^{3}u_{2} d^{3}v_{2} \xi_{2}(\mathbf{u}_{2})\zeta_{2}(\mathbf{v}_{2}) {}_{c}\langle 2|\hat{H} - E_{c}|2\rangle_{c} \chi_{2}(\mathbf{Q}_{2})\xi_{2}(\mathbf{u}_{2})\zeta_{2}(\mathbf{v}_{2}) = 0
\end{eqnarray}

In the first integral equation's diagonal part, $\mathbf{Q}_{1}$,$\mathbf{u}_{1}$ and $\mathbf{v}_{1}$ are linearly independent so we take $\chi_{1}(\mathbf{Q}_{1})$ outside the integral. Since $\chi_{1}(\mathbf{Q}_{1})$ is the only unknown so we can integrate the remaining integrands. In the off-diagonal part of \eqref{A}, we replace $\mathbf{u}_{1}$,$\mathbf{v}_{1}$ with $\mathbf{Q}_{2}$ and $\mathbf{Q}_{3}$ whereas $\mathbf{u}_{2}$,$\mathbf{v}_{2}$ is replaced by a linear combination of $\mathbf{Q}_{1}$,$\mathbf{Q}_{2}$ and $\mathbf{Q}_{3}$. $\mathbf{Q}_{1}$,$\mathbf{Q}_{2}$ and $\mathbf{Q}_{3}$ form a set of linearly independent vectors.
The same steps apply for the integral equation \eqref{B}. After performing some differentiations and integrations, we obtain the following two equations from \eqref{A} and \eqref{B}.
\begin{eqnarray}\label{D}
\left(\frac{3\omega}{2} - \frac{\nabla_{Q_{1}}^{2}}{2m} + 24C_{0}d^{2} - \frac{8}{3}\bar{C} - E_{c} + 4m\right)\chi_{1}(\mathbf{Q}_{1}) \nonumber \\
+ \left(-\frac{1}{2}\right)\bigintss d^{3}Q_{2}d^{3}Q_{3} \textmd{exp} - \left(\frac{Q_{1}^{2}+ Q_{2}^{2}+2Q_{3}^{2}}{2d^{2}}\right) \nonumber \\
\times \left[-\frac{8}{6m(2\pi d^{2})^{3}}h_{1}
+ \frac{32}{9(2\pi d^{2})^{3}}(-2 \bar{C}-4CQ_{3}^{2})- \frac{8(E_{c}-4m)}{3(2\pi d^{2})^{3}} \right]\chi_{2}(\mathbf{Q}_{2}) = 0
\end{eqnarray}
where, written up to accuracy 4,
\begin{equation}
h_{1} = 0.0154 (-7.2550 + Q_{1}^{2} + Q_{2}^{2} + Q_{3}^{2})
\end{equation}
Similarly, we also have
\begin{eqnarray}\label{E}
\left(\frac{3\omega}{2} - \frac{\nabla_{Q_{2}}^{2}}{2m} + 24C_{0}d^{2} - \frac{8}{3}\bar{C} - E_{c} + 4m\right)\chi_{2}(\mathbf{Q}_{2}) \nonumber \\
+ \left(-\frac{1}{2}\right)\bigintss d^{3}Q_{1}d^{3}Q_{3} \textmd{exp} - \left(\frac{Q_{1}^{2}+ Q_{2}^{2}+2Q_{3}^{2}}{2d^{2}}\right) \nonumber \\
\times \left[-\frac{8}{6m(2\pi d^{2})^{3}}h_{1}
+ \frac{32}{9(2\pi d^{2})^{3}}(-2 \bar{C}-4CQ_{3}^{2})- \frac{8(E_{c}-4m)}{3(2\pi d^{2})^{3}} \right]\chi_{1}(\mathbf{Q}_{1}) = 0
\end{eqnarray}
We Fourier transform \eqref{D} w.r.t. $\mathbf{Q}_{1}$ and applying the Born approximation, we use
\begin{equation}
\chi_{2}(\mathbf{Q}_{2}) = \sqrt{\frac{2}{\pi}} e^{i\mathbf{P}_{2}.\mathbf{Q}_{2}}
\end{equation}
inside the integral to get
\begin{eqnarray}\label{F}
\left(\frac{P_{1}^{2}}{2\mu_{12}} + 3\omega - E_{c}\right) \chi_{1}(\mathbf{P}_{1}) \nonumber \\
= \left(-\frac{1}{16 \pi^{5}d^{6}}\right)\left(-\frac{1}{2}\right) \bigintss d^{3}Q_{1}d^{3}Q_{2}d^{3}Q_{3} e^{-\frac{Q_{1}^{2}+ Q_{2}^{2}+2Q_{3}^{2}}{2d^{2}}} \nonumber \\
\times \left[-\frac{8}{6m}h_{1} + \frac{32}{9} \left(-2\bar{C}-4CQ_{3}^{2}\right) - \frac{8(E_{c}-4m)}{3}\right] e^{i(\mathbf{P}_{2}.\mathbf{Q}_{2} + \mathbf{P}_{1}.\mathbf{Q}_{1})}
\end{eqnarray}
Here, $\chi_{1}(\mathbf{P}_{1})$ is the Fourier transform of $\chi_{1}(\mathbf{Q}_{1})$. We write the formal solution of \eqref{F} as

\begin{eqnarray}
\chi_{1}(\textbf{P}_{1}) = \frac{\delta(P_{1}-P_{c}(1))}{P_{c}^{2}(1)} - \frac{1}{\Delta_{1}(P_{1})}\frac{1}{16 \pi^{5}d^{6}}\left(-\frac{1}{2}\right)\bigintss d^{3}Q_{1}d^{3}Q_{2}d^{3}Q_{3} e^{-\frac{Q_{1}^{2}+Q_{2}^{2}+2Q_{3}^{2}}{2d^{2}}} \nonumber \\
\times \left[-\frac{8}{6m}h_{1} + \frac{32}{9}(-2\bar{C}-4CQ_{3}^{2}) - \frac{8(E_{c}-4m)}{3}\right] e^{i(\mathbf{P}_{1}.\mathbf{Q}_{1}+\mathbf{P}_{2}.\mathbf{Q}_{2})}
\end{eqnarray}
with,
\begin{equation*}
\Delta_{1}(P_{1}) = \frac{P_{1}^{2}}{2\mu_{12}} + 3\omega - E_{c} - i\varepsilon
\end{equation*}
If \emph{x}-axis is chosen along $\mathbf{P}_{1}$ and \emph{z}-axis in such a way that the \emph{xz}-plane becomes the plane containing $\mathbf{P}_{1}$ and $\mathbf{P}_{2}$, the aforementioned equation takes the following form
\begin{equation}\label{G}
\chi_{1}(\mathbf{P}_{1}) = \frac{\delta(P_{1}-P_{c}(1))}{P_{c}^{2}(1)} - \frac{1}{\Delta_{1}(P_{1})}F_{1}
\end{equation}
where, in the rectangular coordinates,
\begin{eqnarray}
F_{1} = \frac{1}{16 \pi^{5}d^{6}}\left(-\frac{1}{2}\right)\bigintss dx_{1}dy_{1}dz_{1}dx_{2}dy_{2}dz_{2}dx_{3}dy_{3}dz_{3}
e^{- \frac{\left(x_{1}^{2}+y_{1}^{2}+z_{1}^{2}+x_{2}^{2}+y_{2}^{2}+z_{2}^{2}+2\left(x_{3}^{2}+y_{3}^{2}+z_{3}^{2}\right)\right)}{2d^{2}}} \nonumber \\
\times \left[-\frac{8}{6m}h_{1} + \frac{32}{9}(-2\bar{C}-4C(x_{3}^{2}+y_{3}^{2}+z_{3}^{2})) - \frac{8(E_{c}-4m)}{3}\right] e^{iP(x_{1} + x_{2}\cos\varphi + z_{2}\sin\varphi)}
\end{eqnarray}
Here $\varphi$ is the angle between $\mathbf{P}_{1}$ and $\mathbf{P}_{2}$. As we are considering elastic scattering, so
\begin{equation*}
P_{1} = P_{2} = P
\end{equation*}
From \eqref{G}, we can write the 1,2 element of the T-matrix \cite{EPJ} as follows
\begin{equation*}
T_{12} = 2\mu_{12}\frac{\pi}{2}P_{c}F_{1}
\end{equation*}
where,
\begin{equation}
P_{c} = P_{c}(2) = P_{c}(1) = \sqrt{2\mu_{12}\left(E_{c}-(M_{1}+M_{2})\right)}
\end{equation}
and,
\begin{equation*}
M_{1} = M_{2} = \frac{3\omega}{2}
\end{equation*}
For the total spin averaged cross sections, we make use of the following relation \cite{PDG}
\begin{equation}
\sigma_{ij} = \frac{4\pi}{P_{c}^{2}(j)} \sum_{J} \frac{(2J+1)}{(2s_{1}+1)(2s_{2}+2)}|T_{ij}|^{2}
\end{equation}
where $J$ and $s_{1}$, $s_{2}$ denote the total spin of the two outgoing mesons and spins of the two incoming mesons respectively.
In our situation $J=0$ and $s_{1}=s_{2}=1$. Thus, for  $i=1,j=2$ we obtain
\begin{equation}
\sigma_{12} = \frac{4\pi}{P_{c}^{2}}\frac{1}{9}|T_{12}|^{2}
\end{equation}
Similarly, for $i=2,j=1$
\begin{equation}
\sigma_{21} = \frac{4\pi}{P_{c}^{2}}\frac{1}{9}|T_{21}|^{2}
\end{equation}

\section{Results and Discussion}
\subsection{Results with Quadratic Potential}
To fit the parameters for quadratic potential, we first took the
spin averaging over $E_i$ to obtain the value of $\omega$ and $c$
for the set of mesons $\eta_c(1S)$, $\eta_c(2S)$, $J/\psi(1S)$ and
$J/\psi(2S)$. Here $E_i=(\omega/2)(4n+2l+3)+c$ \cite{Zettili}. By
using the fitted value of $\omega=0.303$ GeV and $c=2.61$ GeV the constant
$C=-(3/16)(2\mu\omega^2)$, mesons sizes $d=\sqrt{1/2\mu\omega}$ and
$\bar{C}=-(3/4)(c-2m)$ are obtained. The constituent quark mass is taken from
ref.\cite{Barnes2005} and the meson mass is obtained from
\cite{PDG}. Hence the required parameters are $d$ = 1.49 GeV
$^{-1}$, quark mass m = 1.48 GeV, $M_{J/\Psi}$ =
3.10 GeV, $C$ = -0.0255 GeV$^{3}$ and $\bar{C}$ = 0.259 GeV.
The obtained results are as under:

\begin{table}[H]
  \centering
  \caption {Total spin averaged cross sections versus selected values of $T_{c}$ for Quadratic Potential}
\begin{tabular}{c c}
\hline\hline
Total COM energy T$_{c}$ (GeV) & Total cross section $\sigma$ (mb) \\ [0.5ex]
\hline
0.01 & 1.33 \\
0.02 & 1.06 \\
0.03 & 0.852 \\
0.1 &  0.158 \\
0.19 & 0.00886 \\
0.2 & 0.00570 \\
\hline
\end{tabular}
\end{table}
The results clearly indicate that with an increase in $T_{c}$ the total cross section gradually decreases.

\subsection{Results with Cornell potential}

For Cornel potential the parameters $\alpha_s$ and $C$ are adjusted by
minimizing the $\chi^{2}$  between the masses taken from
\cite{PDG} and a spectrum generated by using Cornell in the
quark potential model for the mesons $\eta_c$, $J/\psi$, $h_c$,
$\chi_{c0}$, $\chi_{c1}$ and $\chi_{c2}$. The other parameters i.e
string tension, $\sigma$ and quark mass are taken from \cite{Wong Swanson Barnes 2001}.
For the (linear plus Coulombic) potential, the values of the fitted parameters are,
$d$ = 0.995 GeV$^{-1}$, constituent quark mass, $m$ = 1.93 GeV,
$M_{J/\Psi}$ = 3.10 GeV, $\alpha_{s}$ = 0.5, $b_{s}$ =
0.18 GeV$^{2}$. We obtain the following results:

\begin{table}[H]
  \centering
  \caption {Total spin averaged cross sections versus selected values of $T_{c}$ for Cornell Potential}
\begin{tabular}{c c}
\hline\hline
Total COM energy T$_{c}$ (GeV) & Total cross section $\sigma$ (mb) \\ [0.5ex]
\hline
0.02 & 0.15 \\
0.04 & 0.11 \\
0.1 & 0.07 \\
0.2 & 0.03 \\
0.3 & 0.01 \\
0.5 & 0 \\
\hline
\end{tabular}
\end{table}

A graphical comparison of the scattering cross section results obtained through Cornell and Quadratic potentials respectively is shown below.
\begin{figure}[H]
\centering
  \includegraphics[scale=0.15]{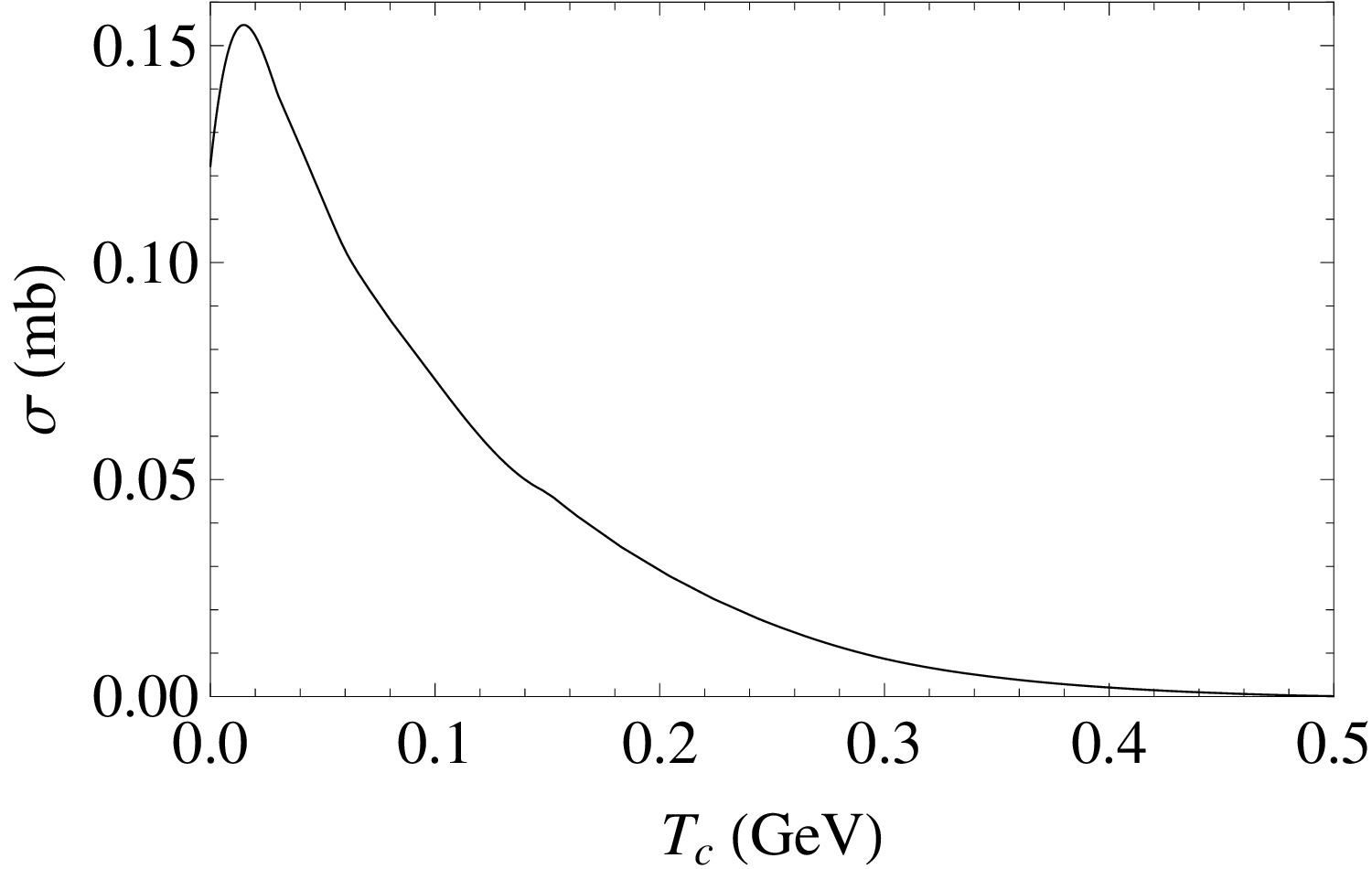}
 \caption{\textbf{Figure 3} \hspace{1mm} Total spin averaged cross sections  versus $T_{c}$ for  Cornell potential}
\end{figure}
\vspace{7mm}
\begin{figure}[H]
\centering
  \includegraphics[scale=0.15]{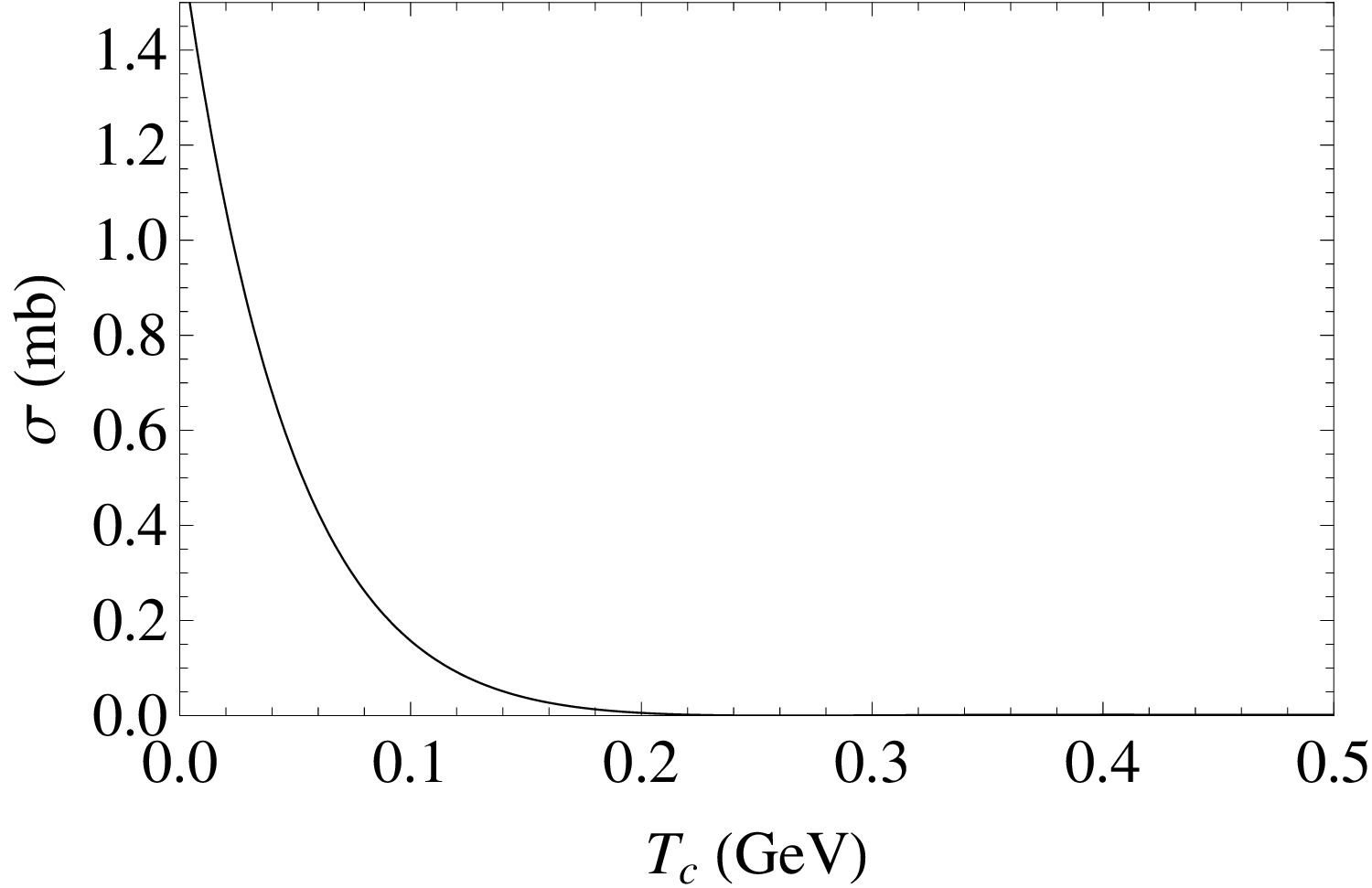}
  \caption{\textbf{Figure 4} \hspace{1mm} Total spin averaged cross sections  versus $T_{c}$ for Quadratic Potential}
\end{figure}

\subsection{Results with Coulombic plus Quadratic Potential}
After a discussion of Cornell and Quadratic potentials respectively, we also incorporate Coulombic plus Quadratic potential in our study. It is defined as
\begin{equation}
v_{ij} = Cr^{2}_{ij} - \frac{4}{3} \frac{\alpha_{s}}{r_{ij}} + \bar{C}
\end{equation}
where $i,j = 1,2,\bar{3},\bar{4}$

For Coulombic plus Quadratic potential the parameters $\alpha_{s}$,$C$ and $\bar{C}$ are adjusted by minimizing the $\chi^{2}$ between the masses taken from \cite{PDG} and a spectrum generated by using the Coulombic plus Quadratic potential in the quark potential model for the mesons $\eta_{c}$ , $J/\Psi$ ,$h_{c}$ ,$\chi_{c0}$ ,$\chi_{c1}$,and $\chi_{c2}$. The quark mass is taken from \cite{Wong Swanson Barnes 2001}. For the Coulombic plus Quadratic Potential, the values of the
parameters are, $d$ $=1.00$ GeV$^{-1}$, constituent quark mass $m$ $= 1.93$ GeV, $M_{J/\Psi}$ $=3.10$ GeV, $\alpha_{s}$ $=0.3$ , $C$ $=-0.047$ GeV$^{3}$ and $\bar{C}$ $=0.792$ GeV. The results are as follows:

\begin{table}[H]
  \centering
  \caption {Total spin averaged cross sections versus selected values of $T_{c}$ for Coulombic plus Quadratic Potential}
\begin{tabular}{c c}
\hline\hline
Total COM energy T$_{c}$ (GeV) & Total cross section $\sigma$ (mb) \\ [0.5ex]
\hline
0.01 & 0.282 \\
0.02 & 0.269 \\
0.03 & 0.244 \\
0.04 & 0.222 \\
0.1 & 0.121 \\
0.2 & 0.0472 \\
0.3 & 0.0173 \\
0.4 & 0.00544 \\
0.5 & 0.00121 \\
\hline
\end{tabular}
\end{table}

\begin{figure}[H]
\centering
  \includegraphics[scale=0.3]{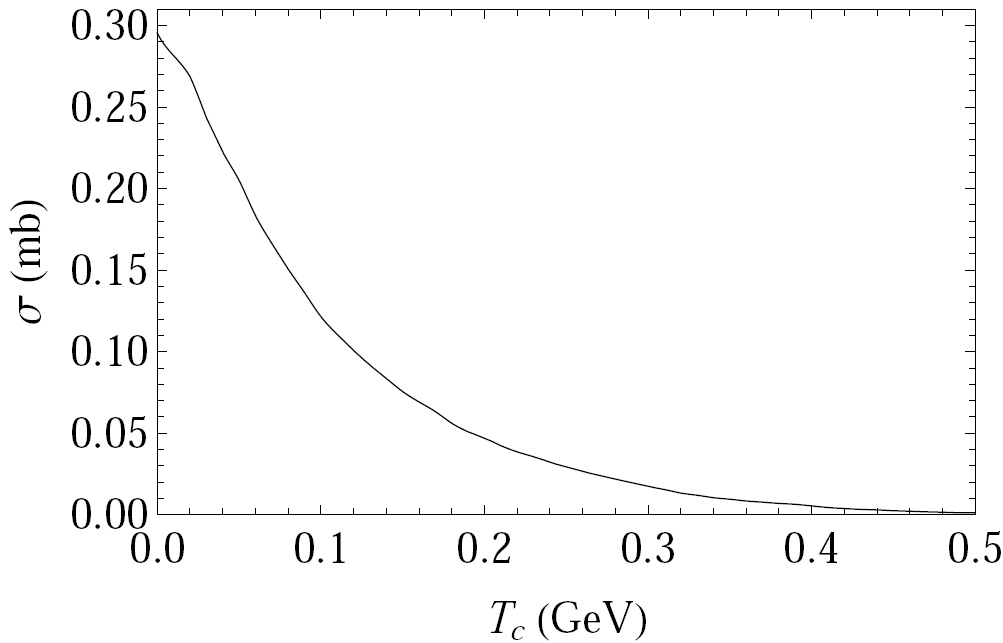}
  \caption{\textbf{Figure 5} \hspace{2mm} Total spin averaged cross sections  versus $T_{c}$ for Coulombic plus Quadratic Potential}
  \end{figure}

 It is observed that if we replace linear confinement with the quadratic in the Coulombic plus linear potential, then the magnitude of the cross sections lie in between the two i.e. the quadratic and the Cornell potentials. As noted in our earlier work \cite{JAM} , the meson sizes obtained via two potential models show a remarkable difference in magnitude i.e. the meson size ‘$d$’ calculated for Quadratic potential is about 1.5 times greater than its value for the Cornell potential. A possible explanation could take into account the respective shapes of the two potentials for any given value of energy $E$. Thus, for a specific energy $E$, the value of classical turning point $r_{0}$ ($E$ $=$ $V(r_{0})$) is greater for Quadratic potential than the Cornell potential and  beyond classical turning point (where $E$ $<$ $V(r)$) the state-function dampens rapidly. This means that a greater value of classical turning point for quadratic potential gives us a larger rms radius compared to the Cornell potential. The rms radii for the Cornell and Quadratic potentials are  $d$ $=$ $0.995$ GeV$^{-1}$ and $d$ $=$ $1.49$ GeV$^{-1}$ (where $d =$ $\sqrt{\frac{1}{2 \mu \omega}}$ \cite{MASUD}) respectively.
\par
Lastly, we can also study the effect of scattering angles on the aforementioned cross sections for quadratic potential. The results for different angles (such as $\varphi$ = $0^{\circ}$,$30^{\circ}$,$60^{\circ}$,$90^{\circ}$) between the two incoming waves can be plotted. However, it can be shown that varying the scattering angle has no effect whatsoever on the respective cross sections i.e. the graphs for different values of scattering angle $\varphi$ overlap if plotted simultaneously.  Hence, the scattering cross sections are independent of the angles between the two incoming waves.

\section{Conclusions}
We have considered J/ $\Psi$ - J/$\Psi$ scattering using Quadratic and Cornell potentials.  The qualitative behavior of both graphs is identical i.e. with an  increase in  total COM energy the scattering cross sections gradually decrease. However, by using quadratic potential we obtain higher magnitudes of scattering cross sections. It is pertinent to mention that throughout our discussion, we have reported our model-dependent results upto three significant figures.
\par
At low energies only S-wave scattering is significant. The contribution to scattering cross sections from other partial waves, I $>$ 0, is negligible.
As we are studying the case of S-wave, so it can be seen that as the energy increases, scattering cross sections decrease which is in accordance with the experimental results (lattice QCD simulations) for other scatterings. The quadratic confinement is a good approximation of more realistic Cornell potential as far as the properties of conventional hadrons are concerned. However, this comparison shows that as far as the tetraquark systems are concerned the quadratic confinement does not appear to be a good approximation of Cornell confinement.
So, in order to find out the properties of four quark systems like the spectra of four quark states, it is recommended not to use the quadratic confinement as a replacement of Cornell potential for future studies.
\par
Since heavy quarkonia decay strongly, it is very difficult to directly measure the dissociation cross sections in hadron scattering experiments. As we mentioned in our introduction, the cross sections are calculated using theoretical approaches. One such theoretical approach called the quark-interchange model \cite{SWA} has been successfully supported with light hadron scattering data ($I$ = 2 $\pi\pi$ \cite{SWA}, $I = 3/2$ $K\pi$ \cite{wswan}, $I = 0,1$ $K,N$ \cite{barnswand}). While we have retained basic features of quark-interchange model, our improvement to certain treatments in it suggest that our calculations may be reasonably sound as far as the heavy hadron-hadron scattering is concerned. However, a direct comparison with experimental data on heavy hadron interaction is not possible. In future, it may be useful to check out our predicted cross sections using detailed Monte-Carlo simulations. If our work turns out to be reasonably accurate, then it will be clearly beneficial to include our insights in simulations involving hadron processes in heavy-ion collisions and other studies concerning tetraquark states.

\end{document}